\documentclass[prb,
twocolumn,
 jmp,%
 amsmath,amssymb,
]{revtex4-1}

\usepackage{graphicx}
\usepackage{dcolumn}
\usepackage{bm}
\usepackage{mathtools}
\usepackage{hyperref}
\usepackage{mathrsfs}

\hypersetup{
    colorlinks,
    citecolor=black,
    filecolor=black,
    linkcolor=black,
    urlcolor=black
}

\newcommand{\norm}[1]{\left\lVert#1\right\rVert}

\begin{document}


\title[]{Equilibrium Field Theory of  Magnetic Monopoles in Degenerate Square Spin Ice: Correlations, Entropic Interactions, and Charge Screening Regimes}

\author{Cristiano Nisoli}

\affiliation{ 
Theoretical Division, Los Alamos National Laboratory, Los Alamos, NM, 87545, USA
}%


\date{\today}

\begin{abstract}
We describe degenerate square spin as an ensemble of magnetic monopoles  coupled via an emergent entropic field that subsumes the effect of the underlying spin vacuum. We compute their effective free energy, entropic interaction, correlations, screening, and structure factors that coincide with the experimental ones. Unlike in pyrochlore ices, a dimensional mismatch between real and entropic interactions leads to weak singularities at the pinch points and algebraic correlations at long distance. This algebraic screening  can be, however, camouflaged by a pseudo-screening regime. 
\end{abstract}

\maketitle

{\it Introduction.} Magnetic monopoles~\cite{ryzhkin2005magnetic,Castelnovo2008} provide an emergent description of  the low energy physics of rare earth spin ices~\cite{Ramirez1999,den2000dipolar,Bramwell2001} and have raised considerable interest, in particular in regard to ``magnetricity'' ~\cite{morris2009dirac,Giblin2011,blundell2012monopoles,kirschner2018proposal}.
Spin ices can be modeled~\cite{den2000dipolar,henley2005power,isakov2004dipolar} as systems of Ising spins on a pyrochlore lattice, impinging on tetrahedra, such that their low energy state obeys the Bernal-Fowler ice rule~\cite{giauque1936entropy,bernal1933theory,Pauling1935}: two spins point in, two out of each tetrahedron, realizing a degenerate manifold of constrained disorder~\cite{henley2010coulomb,castelnovo2010coulomb} and residual entropy~\cite{Pauling1935,Ramirez1999}. Then, violations of the ice rule can be interpreted as sinks or sources of the magnetization, {\em i.e.}\ charges,  and the low energy physics of spin ice can be described in terms of mobile, deconfined, {\em magnetic monopoles}, interacting via a Coulomb law, in a disordered spin vacuum.  
Spin ice is thus a prominent platform in which monopole physics can be investigated. 

While experimental probes of monopoles in crystal-grown spin ices are necessarily indirect and  at low temperature,  magnetic monopoles can now be characterized directly in real time, real space~\cite{ladak2011direct,Mengotti2010, Zhang2013,farhan2019emergent} at desired temperature and fields in {\em artificial spin ices}.~\cite{tanaka2006magnetic,Wang2006,Nisoli2013colloquium,heyderman2013artificial,ortiz2019colloquium} These 2D arrays of magnetic, frustrated nanoislands are fabricated in a variety of geometries, often for exotic behaviors not found in natural magnets.~\cite{Morrison2013,farhan2017nanoscale,bhat2014non,shi2018frustration,nisoli2018frustration,skjaervo2019advances,gliga2019architectural,sklenar2019field}

Here we study the monopoles of degenerate  artificial square ice,~\cite{Moller2006,schanilec2019artificial} recently realized in nanopatterned magnets~\cite{perrin2016extensive,ostman2018interaction,farhan2019emergent} and in a quantum annealer~\cite{king2020quantum}. Square ice provides a direct 2D analogue of 3D pyrochlore spin ice where monopole excitations can be directly characterized.   
Heuristic field theories of rare earth pyrochlores \cite{henley2005power,isakov2004dipolar,henley2010coulomb,castelnovo2010coulomb} have dealt only with the ice manifold, via a coarse grained, solenoidal field that describes the spin texture. 
We propose instead a framework where magnetic monopoles are the constitutive degrees of freedom, while the underlying spin ensemble is subsumed into entropic forces among these topological defects. 

We show that, in absence of physical interaction~\cite{king2020quantum}, monopoles interact {\em entropically} via a 2D-Coulomb, logarithmic law, leading to $(2+1)$ electromagnetism, Bessel correlations of finite screening length for $T>0$, and thus a conductive phase~\cite{berezinskii1971destruction,kosterlitz1973ordering} where monopoles are unbound. The correlation length diverges exponentially as $T\downarrow 0$, signaling the criticality  of the ice manifold.  
Then, inclusion of the monopole-monopole 3D-Coulomb interaction, leads, unlike in   3D spin ice,  to algebraic correlations and weak singularities in the structure factor at $T\ne 0$. Its interplay with the entropic interaction drives screening regimes of  {\em effectively} bound and unbound monopoles. 

\begin{figure}[b!]
\includegraphics[width=.9\columnwidth]{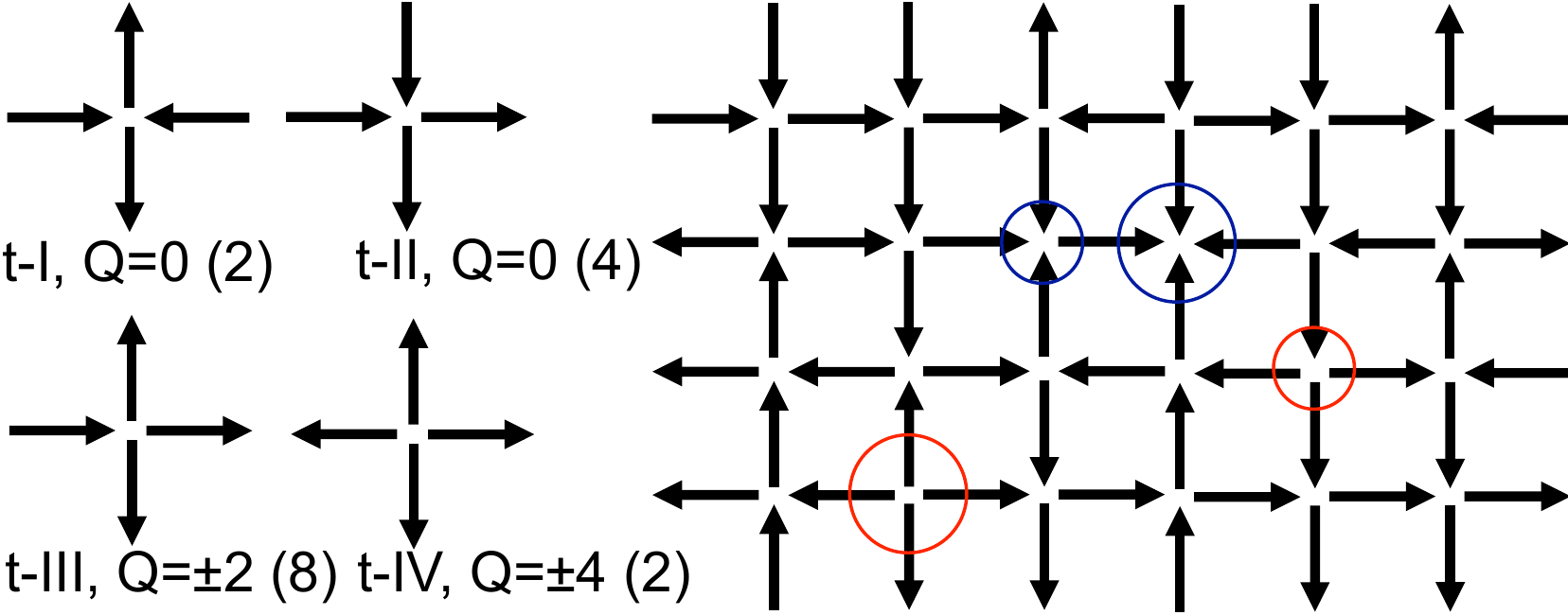}
\caption{Left: the four vertex configurations, ice-rule ones at the top ($Q=0$) and monopole at the bottom (multiplicities in parenthesis). Right: portion of spin ice with monopoles of different charge circled.}
\label{fig1_0}
\end{figure}

{\it 1. Field Theory.} Square ice is a set of $N_e$ classical, binary spins $\vec S_e$ aligned on the edges $e$ of a square lattice of unit vectors $\hat e_1, \hat e_2$, of $N_v=N_e/2$ vertices labeled by $v$. The 16 spin configurations of a vertex are classified by 4 topologies~\cite{Wang2006, perrin2016extensive} as t-I, \dots, t-IV  (Fig.~1). 
Of each vertex, the charge $Q_v=2n-4$ is equal to the number $n$ of spins pointing in minus the $4-n$ pointing out.
Then,  $Q=0$ for ice-rule obeying vertices (t-I, t-II). 
The following    
\begin{align}
{\cal H}[Q]=\frac{\epsilon}{2}\sum_v Q_v^2  + \frac{\mu}{2}\sum_{v \neq v'} Q_vV_{vv'} Q_{v'}, 
%
\label{H-n}
\end{align}
is a suitable Hamiltonian for a variety of square ice realizations: 
 $\epsilon$ is the cost of monopoles and  $V_{vv'}=1/ 2\pi |v-v'|$ is their 3D-Coulomb interaction; we take the lattice constant  $a=1$ and thus $\mu$ is an energy. When $\mu/\epsilon < 2\pi/M_{\square}\simeq 3.9$  (where $M_{\square}\simeq1.6155$ is our Madelung constant~\cite{borwein1985convergence}) the ground state is a disordered  tessellation of the six ice rule vertices (Fig.~1) of known Pauling entropy~\cite{lieb1967residual,Baxter1982}. Above that threshold, the ground state becomes unstable toward the formation of a monopole ionic crystal.

In magnetic realizations,~\cite{perrin2016extensive,ostman2018interaction,farhan2019emergent} Eq.~(\ref{H-n}) corresponds to the dumbbell model of ref~\cite{Castelnovo2008} (then $\epsilon$ is the coupling of the dumbbell charges in the vertex) with the  3D-Coulomb interaction comes from the usual truncated multipole expansion, and  therefore $\mu = {\mu_0 p^2}/{ l_d^2}$, with $p$ the magnetic moment, $l_d<a$ the dumbbell length. Then, $\mu/\epsilon\simeq 1-l_d/a<1$. 
Equation~(\ref{H-n}) does not describe the standard, non-degenerate square ice~\cite{Wang2006,Nisoli2010,Morgan2010,Zhang2013,Porro2013,sendetskyi2019continuous} in which monopoles are confined by the tension of magnetic Faraday lines.\cite{mol2009magnetic,silva2012thermodynamics,silva2013nambu,levis2013thermal, nisoli2020topological,foini2013static} Finally, even in realizations where ice-rule vertices are degenerate~\cite{Moller2006,perrin2016extensive,ostman2018interaction,schanilec2019artificial,king2020quantum}, the long ranged dipolar interaction  favors antiferromagnetic ordering~\cite{perrin2016extensive} at  $T\ll \epsilon$, an effect that we  consider here elsewhere. 

The  partition function is 
\begin{equation}
Z \left[H\right]=\sum_{S} \exp{\! \left(-\beta {\cal H} +\beta \sum_e \vec S_e\cdot \vec H_e\right)},
\label{Z-n0}
\end{equation}
  ($\beta=1/T$) such that $\langle \vec S_{e_1} \dots \vec S_{e_n}\rangle=\partial_{\beta \vec H_{e_1} \dots \beta \vec H_{e_n}} \! \! \ln Z$.
To sum over the spins, we insert the tautology~\cite{hubbard1959calculation}   $1=\prod_v \int   dq_v d\phi_v\exp\left[i \phi_v\! \! \left(q_v-Q_v\right) \right]/(2\pi)^{N_v}$ obtaining
%
%
%
\begin{equation}
Z\left[H\right]=\int \left[dq \right]  e^{-\beta{\cal H}[q]} \tilde \Omega[q]
\label{Z-n}
\end{equation}
where $\left[dq\right]= \prod_v dq_v/(2\pi)^{N_v} $, and  the density of states
\begin{equation}
\tilde \Omega[q]=\int [d\phi] \Omega[\phi]e^{i \sum_v  q_v \phi_v},
\label{Omegatilde-n}
\end{equation}
is the  Fourier transform of the partition function for $\phi$
\begin{equation}
\Omega[\phi]=2^{N_e}{\prod_{{\langle vv' \rangle}}}\cosh \left(-i\nabla_{vv'}\phi + \beta H_{vv' } \right)
\label{Omega-n}
\end{equation}
%
(edges ${\langle vv' \rangle}$ are counted once,  $\nabla_{vv'}\phi \vcentcolon=  \phi_{v'}- \phi_{v}$, $ H_{vv' } \vcentcolon=  \vec H_e\cdot \hat{vv'}$). By construction, $\langle Q_{v_1} \dots Q_{v_n}\rangle=\langle q_{v_1}\dots  q_{v_n}\rangle$. 

We have obtained a theory of continuous charges  constrained by an ``entropy'' 
%
$S[q]=\ln \tilde \Omega[q]$
%
 conveying  the effect of the spin ensemble. 
 Equivalently,  monopoles are  coupled to an  {\em entropic} field~\cite{nisoli2014dumping} $V_{e} = i T \phi$, of ``free energy"  
 \begin{equation}
 {\cal F}[\phi]=-T\ln \Omega[\phi].
 \end{equation}  
From Eqs.~(\ref{Z-n}-\ref{Omega-n}) we have
  \begin{equation}
 \langle S_{vv'}\rangle=\langle \tanh \left(\beta H_{vv'}- i\nabla_{vv'}\phi \right) \rangle,
 \label{M}
 \end{equation}
implying that while $V_e=iT\phi$ correlates charges, $B_e=iT \nabla\phi$ correlates spins that would be  trivially paramagnetic in its absence.
Note that   $\overline {V_e}_v \vcentcolon= \langle V_{e,v} \rangle =i T \langle \phi_v \rangle$ is {\it real}~\footnote{we denote one-point functions with an overline}. In fact,  standard Gaussian gymnastic~\cite{SI} shows  
\begin{equation}
\begin{dcases}
\overline{V_e}_{,v} =\epsilon \overline q_v +  \frac{\mu}{2 \pi}\sum_{v'\ne v} \frac {\overline q_{v'}}{|v-v'|} \\
\overline  q= \sum_{\alpha=x,y}\partial^{\alpha}  \overline{\tanh \left(\beta \partial_{\alpha} {V_e} -\beta H_{\alpha}\right)},
\end{dcases}
\label{eos0}
\end{equation}
with the second equation following from the definition of  $Q$ and from Eq.~(\ref{M}) in the continuum limit. 

{\it 2. Approximations.}    
Eqs.~(\ref{eos0}) allow to compute the charge under boundary conditions in various approximations. 
For $\mu=0$ their linearization  returns  
 screened-Poisson equations  for $q, V_e$,  of screening length
\begin{equation}
\xi_{0}=\sqrt{\epsilon/T}.
\label{xi0}
\end{equation}
(A similar screening length had been found in different geometries via other methods~\cite{garanin1999classical,henley2005power}, and, we show elsewhere, holds for a general graph~\cite{nisoli2020concept}.) 
This approximation corresponds to a high $T$ limit. From  Eqs.~(\ref{Z-n},\ref{Omegatilde-n}), by integrating over $q_v$ one obtains $\langle \phi^2 \rangle \simeq \epsilon/T$: as $T$ increases the system loses correlation and  the entropic field decreases. 

We can therefore legittimately  expand ${\cal F}[\phi]$ in small $\phi$. Then,  Fourier transforming on the Brillouin Zone (BZ)~\footnote{$g(\vec r)=\int_{\text{BZ}}  \tilde g(\vec k)e^{-i \vec k \cdot \vec r }{d^2k}/{(2\pi)^2}$, with $g$ a general function, $\vec r=v,e$  vertices or edges.}, we obtain the approximate partition function 
\begin{align}
Z_{\text{eff}}=\int [dqd\phi]\exp\left({\! -\int_{\text{BZ}} \beta {\cal F}_{\text{eff}}[ q,\phi]( k) \frac{d^2k}{(2\pi)^2}}\right),
\label{Z2-n}
\end{align}
of free energy functional at second order 
\begin{align}
{\cal F}_{\text{eff}}[ q,\phi] & = \frac{\epsilon+\mu \tilde V}{2} \left|\tilde{q}\right|^2 + \frac{T}{2}  \gamma^2  | \tilde   \phi |^2 - iT \tilde{q}^*\tilde{\phi} \nonumber \\
&  -    \tilde{\phi}^*  \vec \gamma \cdot \vec{\tilde{H}}   - \frac{\beta}{2}  \left| \vec{\tilde{H}}\right|^2,
\label{f2-n}
\end{align}
where   $\gamma_{\alpha} (\vec k) \vcentcolon= 2\sin(k_{\alpha}/2)$ and
$\tilde V(\vec k)$ is the Fourier transform of $V$ on the lattice. 
Integrating $Z_{\text{eff}}$ over $\tilde \phi$  returns the {\em effective free energy for monopoles} 
\begin{align}
{\cal F}_{\text{eff}} [q] = \frac{1}{2}\left( \epsilon + \mu \tilde V  +\frac{T}{\gamma^2} \right) |\tilde{q}|^2.
\label{f2rho}
\end{align}
%
The last term implies an entropic interaction among  charges that at large distances ($ \gamma^2 \simeq  k^2$) is
\begin{equation}
{V_e(\vec r_1- \vec r_2)\simeq - T \frac{q_{1} q_{2}}{2\pi }\ln \norm{\vec r_1-\vec r_2}},
\label{Ve}
\end{equation}
i.e.\ 2D-Coulomb. Instead, in 3D, from Eq.~(\ref{f2rho}) the entropic interaction would be 3D-Coulomb, or $\sim 1/r$, thus merely altering the coupling constant $\mu \to \mu +T $ of the real interaction, as  indeed found numerically~\cite{castelnovo2011debye,chern2014realizing}. A charge assignation changes the degeneracy of the spin configurations compatible with it and thus the entropy. That the change in entropy can be written as the sum of pairwise logarithmic interactions  is  not obvious.

Equation~(\ref{f2rho}) implies the charge correlations in $k$ space
\begin{align}
\langle | \tilde  q(\vec {k}) |^2\rangle = \gamma(\vec k)^2 \tilde \chi_{||}(\vec k)
\label{rhocorr}
\end{align}
with $\tilde \chi_{||}(\vec k)$  given by
\begin{align}
\tilde \chi_{||}(\vec k)^{-1}&={1 + {\xi_{0}}^2 \gamma(\vec k)^2\left[1+ \frac{\mu}{\epsilon}\tilde V(k) \right]} \label{taus}.
\end{align}
Note that 
$\langle | \tilde  q(\vec {k}) |^2\rangle$  peaks on the $K$ points of the BZ. 
These peaks diverges when $\mu \uparrow (2\pi/M_{\square})(\epsilon+T/8)$, signaling  the aforementioned instability toward an ionic crystal of $\pm 4$ monopoles.~\footnote{Such large $\mu$ values are unattainable in a  dumbbell model, thus they are hard to realize in magnetic spin ice.} Note also that $\langle q^2\rangle=\int_{\text{BZ}} \langle | \tilde  q(\vec {k}) |^2\rangle {d^2k}/{(2\pi)^2}\uparrow 4$ as $T\uparrow \infty$, which is correct, as it can be verified by considering only vertex multiplicities ($2^2/2+4^2/8=4$).

%

\begin{figure}[t!]
\includegraphics[width=1\columnwidth]{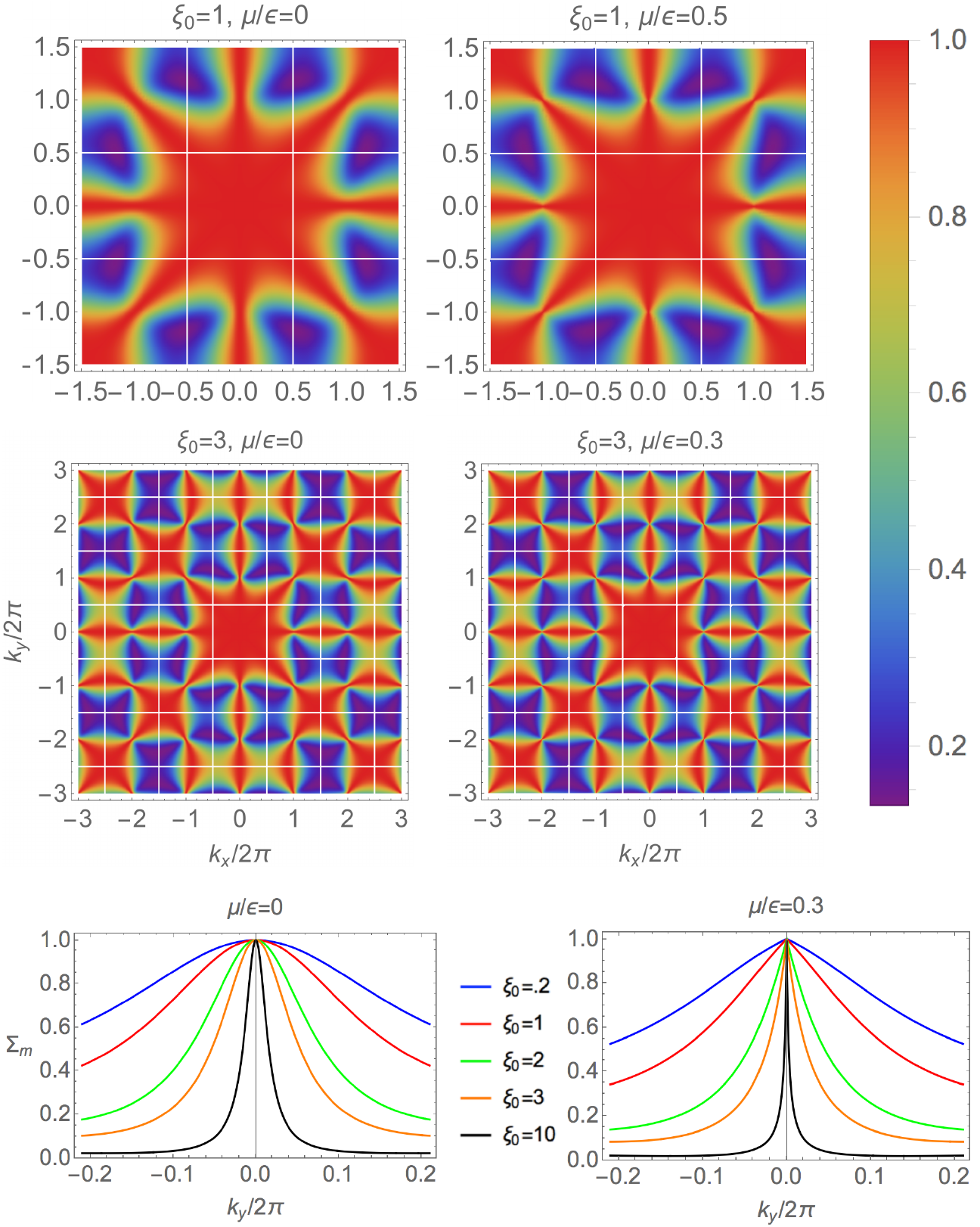}
\caption{Plots of structure factor $\Sigma_m$  obtained from Eq.~(\ref{spincorr}) for $\xi_0=0,3, \mu/\epsilon=0,0.3,0.5$ ($k_x, k_y$ in units of $1/a$); $\mu>0$ leads to sharper pinch points even at high $T$. Bottom row: Structure factors cuts on the line $k_x=2\pi$ demonstrate discontinuity in the first derivative of the intensity when $\mu>0$.}
\label{sf}
\end{figure}

 By performing the integral in Eq.~(\ref{Z2-n}) we obtain
\begin{align}
 \ln Z_{\text{eff}} =  \frac{1}{2} (\beta \vec{\tilde H}^*)\cdot \left( {\hat \gamma \hat \gamma}{ \tilde \chi_{||}} +  { {^{\perp} \! \hat \gamma} {^{\perp} \! \hat \gamma}}\right)\cdot (\beta \vec{\tilde H})
\label{ftot}
\end{align}
($\hat \gamma  \vcentcolon=\vec \gamma/\gamma$,  $^{\perp} \! \hat{\gamma}  \vcentcolon=\hat e_3 \wedge \hat \gamma$), showing that $ \tilde \chi_{||}(\vec k)$  is the  {\em longitudinal  susceptibility} (multiplied by $T$). Thus, 
\begin{equation}
\langle \tilde S^*_{\alpha}(\vec k) \tilde S_{\alpha'}(\vec k)\rangle=  \hat \gamma_{\alpha} \hat \gamma_{\alpha'} \tilde \chi_{||}+   {^{\perp} \! \hat \gamma_{\alpha}} \! \!{^{\perp} \! \hat \gamma_{\alpha'}},
\label{spincorr}
\end{equation}
are the spin correlations, whose   structure factor $\Sigma_m(\vec k)=^{\perp}\! \!\vec{k}\cdot \langle \vec{\tilde{S}}(\vec k) \vec{\tilde{S}}(\vec k) \rangle \cdot ^{\perp}\! \!\vec{k}$  we plot in Fig.~\ref{sf}. In the limit $T\downarrow 0$ and thus  $\xi_0 \uparrow \infty$, correlations in Eq.~(\ref{spincorr}) become  purely transversal. 
When $T>0$, $\mu=0$, pinch points are smoothened by a Lorentzian of width   $\xi_0$. Instead, for {$\mu\ne 0$} the profile is sharper, with weak singularities  controlled by the Bjerrum length $2l_B \vcentcolon =\mu/T$,
\begin{equation}
\Sigma_m(2\pi,k_y)\simeq 1- 2l_b |k_y|~~~\text{at}~~~k_y\simeq 0,
\end{equation}
due to the aforementioned  dimensional mismatch. 

To gain insight on how to proceed at  low $T$,\footnote{Low but ``not too low''. At really low $T$ ordering can take place, as discussed, but also glassy kinetics and non-equilibration.} one  can  perturbatively  expand ${\cal F}[\phi]$. Instead, we  make the reasonable  ansatz that the effective theory  has the same functional form as ${\cal F}_{\text{eff}}$ in Eq.~(\ref{f2rho}) but with constants ``dressed'' by the interactions among fluctuations  at low $T$.  
Note that $\xi_{0} \uparrow \infty $ as $T \downarrow 0$ and  Eq.~(\ref{rhocorr})  implies 
\begin{equation}
\xi_{0} \simeq1/ \sqrt{\langle q^2\rangle} ~~~\text{for} ~T\downarrow 0,
\label{DH}
\end{equation}
and therefore $\epsilon$ is dressed as
%
 $\epsilon \to \epsilon(T) \sim T/\langle q^2\rangle$.
 %
Then, if we approximate $\langle q^2\rangle$  by assuming uncorrelated vertices, we obtain   $ \xi_{0}\simeq ~\! \exp\left({{\epsilon}/{T}}\right)$. This exponential divergence of the correlation length (consistent with experimental findings~\cite{fennell2009magnetic}) points to the topological nature of the critical  ice manifold at $T=0$. Note that
 $ \xi_{0}$ in Eq.~(\ref{DH}) is exactly the Debye-H\"uckel~\cite{levin2002electrostatic}  length for a Coulomb potential of coupling constant proportional to $T$, as is the case of the entropic potential. In Supp. Mat.~\cite{SI},  a  Debye-H\"uckel ~\cite{castelnovo2011debye,kaiser2018emergent} 
 approach inclusive of  the entropic field~\cite{nisoli2014dumping} leads to the very  Eq.~(\ref{rhocorr}) {\em yet} with $\xi_0$ given by Eq.~(\ref{DH}), corroborating our ansatz. 


When $\mu=0$, Eq.~(\ref{f2rho}) reduces to a 2D Coulomb gas and from Eq.~(\ref{rhocorr},\ref{taus})  the charge correlations at large distance are
\begin{align}
&\langle q_{\vec r_1} q_{\vec r_2}\rangle \simeq -\frac{1}{2\pi \xi_{0}^{4}} K_0\left(\norm{\vec r_1-\vec r_2}/{\xi_{0}}\right),
\label{K0charge}
\end{align} 
as recently experimentally verified~\cite{king2020quantum}, 
 showing that $\xi_{0}$  is indeed the correlation length.
 $\xi_{0}$ is also the screening length: a charge $Q_{\mathrm{pin}}$ pinned in $v_0$ elicits a charge   
\begin{equation}
\overline q_v =Q_{\mathrm{pin}} {\langle q_v q_{v_0}\rangle}/{\langle  q^2\rangle}.
\label{qpin}
\end{equation}
%
A finite screening length implies that the system is conductive. There is no BKT transition~\cite{berezinskii1971destruction,kosterlitz1973ordering} to an insulating phase (i.e\ algebraic correlations, bound charges) because the interaction among charges is purely entropic, has coupling constant proportional to $T$, and thus no  interplay between entropy and energy can  drive a transition. The lack of such transition can be shown in general from the model, regardless of our formalism.~\footnote{Nonetheless, it is possible to show that the state at $T=0$ does belong to the critical portion of the KT transition, and indeed the correlation length is infinite.} Yet, when $\mu>0$ the system is always  insulating, as we show now.

{\it 3. Monopole Interactions and Algebraic Correlations}. Consider now $\mu>0$. At  small $k$, 
 $\tilde V(k) \sim 1/k$  and  Eq.~(\ref{qpin}) reads   
\begin{equation}
\overline{ \tilde q} (k)\simeq \frac{k^2}{1 + 2 l_B k + \xi_0^2 k^2 }\frac{Q_{\text{pin}}}{\langle q^2 \rangle},
\label{cincio}
\end{equation}
 which is not analytical at $k=0$, leading to the aforementioned  weak singularities  at the pinch points. 
In  3D  it would be  analytical, because $\tilde V(k) \sim~1/k^{2}$, the poles of $\overline{ \tilde q} (k)$ would be purely imaginary [$k_{\pm} = \pm i / \xi_{\text{3D}}$ with $\xi_{\text{3D}}^{-2}=\xi_{0}^{-2}(1+\mu/T)$], and thus  $\xi_{\text{3D}}$ would be a screening length.  
But in 2D the poles are $k_{\pm} = -\bar k \pm i / \xi_{\mu}$  with
\begin{equation}
\frac{\xi^2_{\mu}}{\xi_0^2}=\frac{\xi^2_{0}}{\xi^2_{0}-{l_B^2}} =\frac{T}{T-T_{\times}}
\end{equation}
and they always  have a real part $\bar k =l_B/ \xi^2_{0}=\mu/2\epsilon$.

  Above the  crossover temperature 
 $T_{\times}=\mu^2/4\epsilon$, $\xi_0 > l_B$, $\xi_{\mu}$ is real, and poles have imaginary parts $i / \xi_{\mu}$. Thus, one could  heuristically consider $\xi_{\mu}$ a screening length, and above $T_{\times}$ monopoles are unbound, and the phase is conducive. Below $T_{\times}$,  $\xi_{\mu}$ is imaginary,  and one could say that there is no screening length, and monopoles are bound by the strength of the magnetic interaction ($\xi_0 < l_B$), in an insulating phase. However, things are more complicated than this naive picture.

In fact, mathematically speaking, there is {\em never} a finite screening length. To demonstrate it, consider a  charge $Q_{\text{pin}}$ pinned in the origin.  From Eqs.~(\ref{f2rho},\ref{Ve}), 
 $ \overline {\tilde V_e}(k)  =T \overline{ \tilde q} (k)/ k^{2}$, and thus we obtain
\begin{equation}
\beta \overline {\tilde V_e}(k)  = \frac{\xi_{\mu}}{2 i \xi_0^2}\left(\frac{1}{k-k_+}-\frac{1}{k-k_-}\right) Q_{\text{pin}}.
\label{Vediv}
\end{equation}
Using  $2/c=\int_{-\infty}^{\infty}\! \exp(-|z| c) dz$ for $\Re(c)>0$ on each fraction in Eq.~(\ref{Vediv}) and Fourier transforming, we  have
\begin{equation}
\beta \overline {V_e}(r)   = \frac{ l_B }{2\pi \langle  q^2 \rangle} \int_{-\infty}^{+\infty} \! \! \! \frac{\lambda(z)}{(r^2+z^2)^{3/2}}dz
\label{Ve2}
\end{equation}
where $\lambda(z)$ can be interpreted as a linear charge density 
\begin{equation}
\lambda(z)=\frac{\xi_{\mu} }{2 \xi_0^2 l_B}  |z| {\sin\left(|z|/\xi_{\mu}\right)}e^{-\bar k |z| }Q_{\text{pin}},
\label{lambda}
\end{equation}
for which  $\int \lambda(z) dz=Q_{\text{pin}}$. 
Therefore,
 the entropic potential  can be represented as if generated by a {\em image} charge~\cite{jackson2007classical},  spread along a line (of coordinate $z$) perpendicular to the 2D system, and of total charge $Q_{\text{pin}}$. 

\begin{figure}[t!]
\includegraphics[width=.999999\columnwidth]{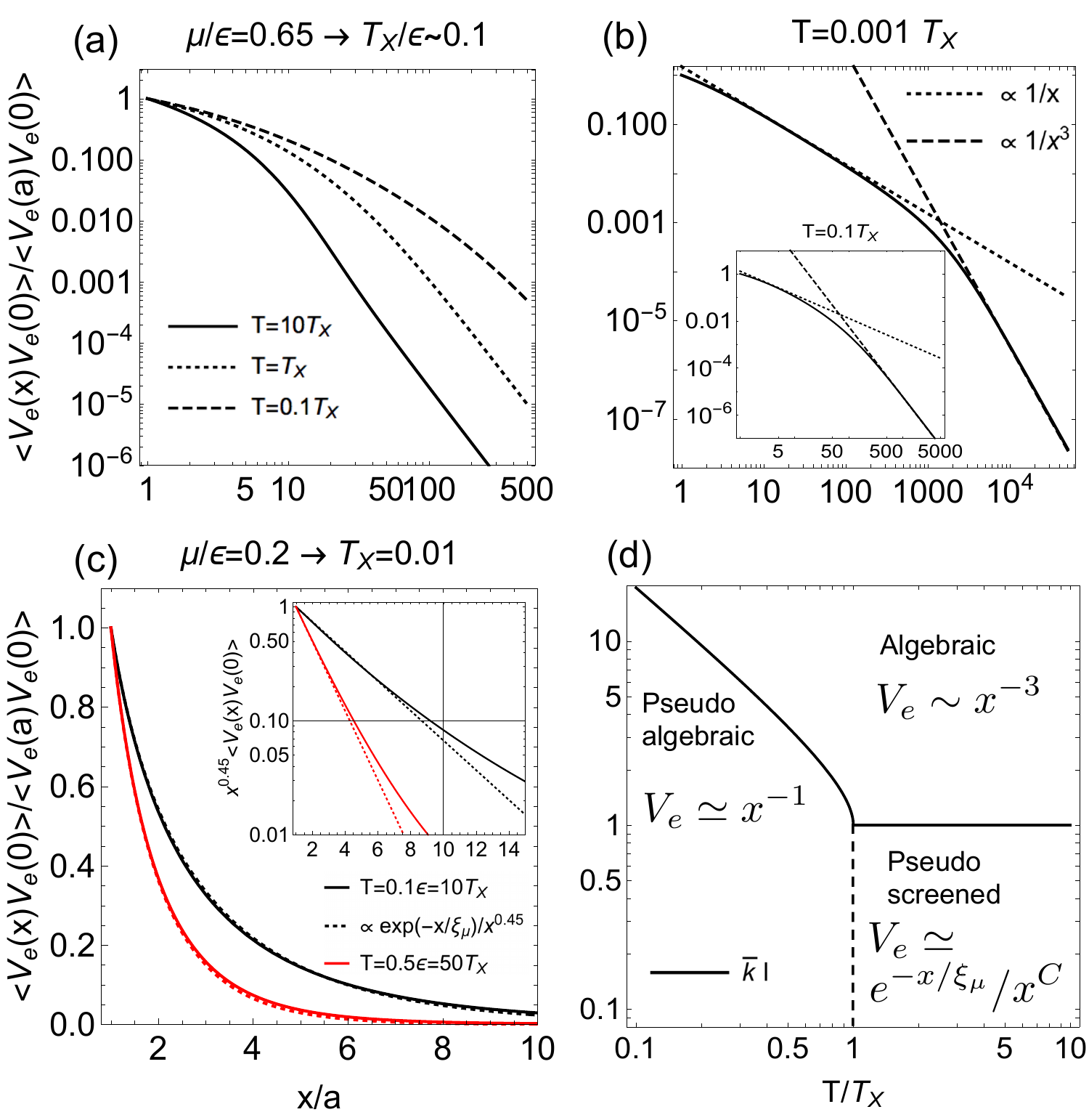}
\caption{Screening behavior at different $T,\mu$. (a) Log-log plots of the screening entropic potential $\overline V_e(x)$ (numerically integrated) of Eq.~(\ref{Ve2}) for different temperatures at relatively strong monopole interaction $\mu =0.65 \epsilon$ leading to $l\simeq 3$ and $T_{\times}=0.1\epsilon$, where our approximation should still apply. Note the algebraic $~1/r^3$ decay. However, at high $T$ the potential drops by 99\% before becoming algebraic. (b) For  $\mu =0.65$ and $T/T_{\times}=10^{-3}$,  $\overline V_e(x)$  shows a pseudo-algebraic decay $\sim 1/r$ for most of its measurable tail (in inset a higher T case,  $T/T_{\times}=0.1$). (c) At low monopole interaction ($\mu=0.2 \epsilon$) $T_{\times}/\epsilon= 0.01$ is very low, most of the charge is screened before the algebraic regime ($l=10$). Effectively, the screening is exponential, and $V_e(x)\propto e^{-x/\xi_{\mu}}/x^{0.45}$ provides a good fit (in inset, log plot of $x^{0.45}V_e(x)$). (d) Plot of $\bar k l$ as a function of $T$ and schematics of  the screening at different distances.}
\label{screen}
\end{figure}

Crucially, $\lambda(z)$ is exponentially confined by a length $l$.
When $T>T_{\times}$,  $l=1/\bar k$. When $T<T_{\times}$, the sine in Eq.~(\ref{lambda}) becomes hyperbolic and  
$l=1/  k_+$. For $r \gg l$ the charge is seen as point-like, the  potential scales as
\begin{equation}
\overline V_e(r)   \simeq \frac{ l_B }{2\pi \langle  q^2 \rangle} \frac{Q_{\text{pin}}}{r^3},
\label{Ve3}
\end{equation}
and, by taking its Laplacian,  $\overline q(r)$ scales as 
\begin{equation}
\overline q(r)   \simeq - \frac{ 9 l_B }{2\pi \langle  q^2 \rangle} \frac{Q_{\text{pin}}}{r^5}.
\label{rhopin}
\end{equation}

Remarkably,  {\em long distance screening---and thus spin correlations---are algebraic} at any $T$ (Fig~\ref{screen}a,b). In 3D spin ice, instead, spin correlations are algebraic only at $T=0$, even with  monopole interaction. 
Algebraic screening from a 3D-Coulomb potential in the polarizability of quantum or classical 2D charge systems has been rediscovered multiple times~\cite{ando1982electronic, keldysh1979coulomb,jena2007enhancement,cudazzo2011dielectric}. In fact, it is not merely a 2D feature, but can happen in any dimension for the same dimensional mismatch in the Coulomb interaction~\cite{Nisoli2020Coulomb}. 
\begin{figure}[t!]
\includegraphics[width=.9\columnwidth]{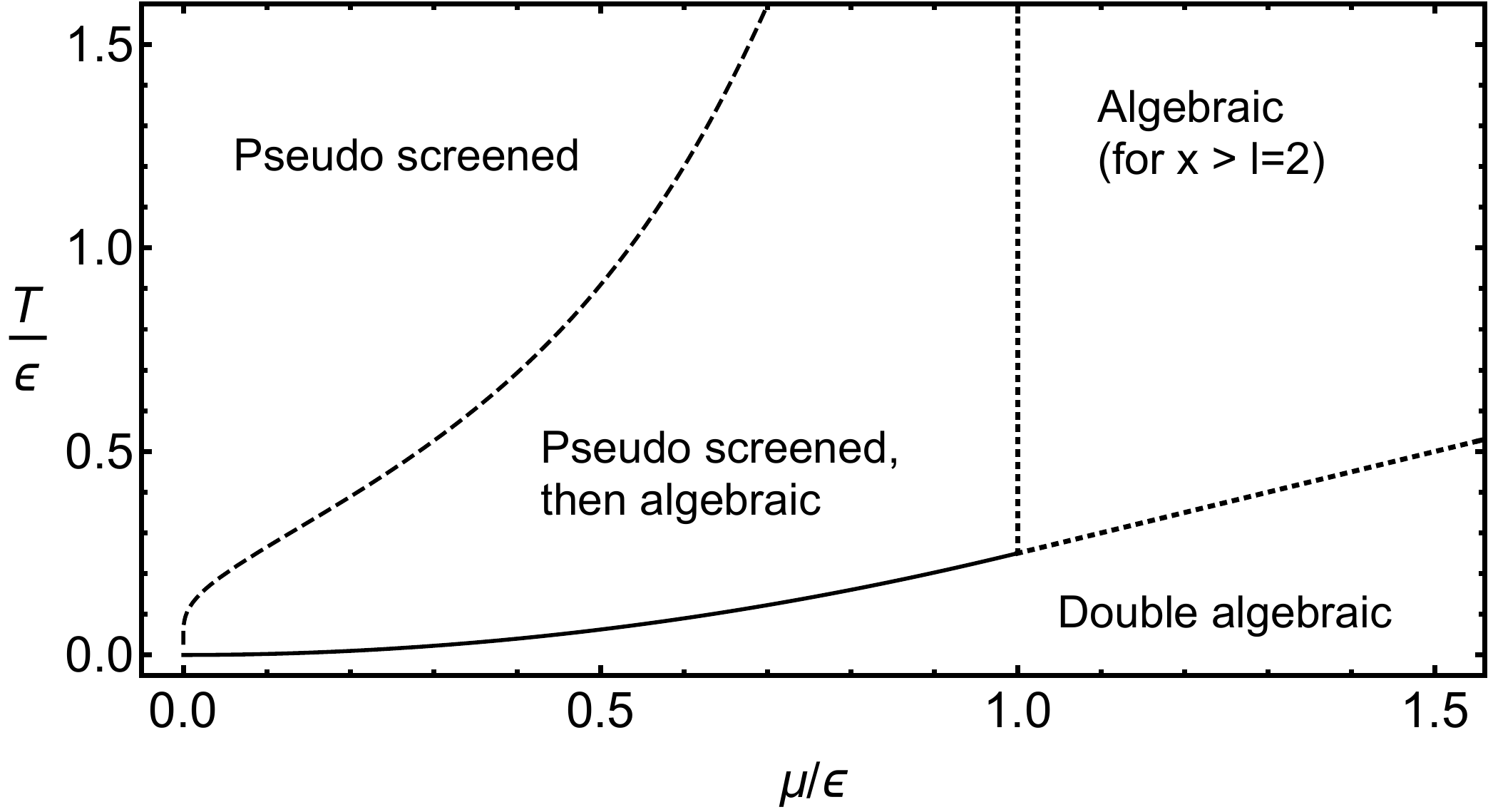}

\caption{Heuristic regime diagram. The solid line is $T_{\times}(\mu)$. The dashed line corresponds to 99\% of the charge being screened within the radius $l$. On the right of the dotted line, $l \le 2$ and there is therefore  {\em algebraic} screening, $\left[\langle q(r)q(0)\rangle\sim r^{-5}\right]$. On the left of the dashed line 99\% of the charge is screened within $l$, and, because $T>T_{\times}(\mu)$, the regime is   {\em pseudo screened}. In the {\em pseudo screened, then algebraic} regime the behavior is effectively screened for $x <l$ but more than 1\% of the charge is algebraically screened  at $x>l$. In the {\em double algebraic} region  $\langle q(r)q(0)\rangle\sim r^{-3}$ for $x\ll l$ and   $\langle q(r)q(0)\rangle\sim r^{-5}$ for $x \gg l$. Note that in nanomagnetic realizations the dumbbell model imposes $\mu/\epsilon <1$.}
\label{pseudo}
\end{figure}
Importantly, unlike electrical charges, magnetic monopoles are emergent particles of a spin ensemble, and interact  entropically. The interplay between  the correlation length at zero interaction ($\xi_0$) and the Bjerrum length ($l_B$) leads to a crossover at $T_{\times}$ between {\it  effectively} conductive and insulating regimes.
 
 To see that, consider 
\begin{equation}
Q_{\text{alg}}/Q_{\text{pin}}\simeq-  3 l_B /\langle  q^2 \rangle l^3,
\end{equation}
the fraction of the charge screened algebraically, obtained by  integrating Eq.~(\ref{rhopin}) for $x>l$. When  it is very small, and $l$ is large, the algebraic nature of the screening might not be  detectable.   

When $ T\gg  T_{\times} $, $|Q_{\text{alg}}/Q_{\text{pin}}| \simeq10^{-2}$ or less, and thus the algebraic screening might not be experimentally detectable. Moreover, as Fig~\ref{screen}c shows,  the screening within $l$ is well fitted by an exponentially screened function. Above $T_{\times}$ there is  a  ``pseudo screening length'' $\xi_{\mu}>l_B$.

When $T \ll T_{\times}$ we can take $k_+\sim0$ and from Eq.~(\ref{Vediv})  $\overline{\tilde V_e}(k)  \sim1/k$, and thus $\overline  V_e(r)  \sim 1/r$ for $r \ll l=1/k_+$ (as confirmed by the numerical plot in Fig~\ref{screen}b). In this regime the insulating phase is easily detectable. 

Considering $l$, $T_{\times}$, and $Q_{\text{alg}}$, we can sketch heuristic regime diagrams, which are necessarily somehow arbitrary as they depend on practical specifications. Clearly, when $l$ is smaller than, say, $2$  the behavior is completely algebraic. When $l$ is instead large, there might be pseudo-screening for $x<l$ if $T>T_{\times}$ and if most (we choose 99\% in figure) of the charge is screened within a radius $l$ (left side of the dashed line in Fig.~\ref{pseudo}).  If $l\gg2$ and $Q_{\text{alg}}/Q_{\text{pin}}$ is not small, an initial exponential screening for $r<l$ is followed by algebraic screening for $r \gg l$. 
Finally, note that $\epsilon$ gets dressed at low temperature. At low $\mu/\epsilon$, the algebraic screening might be extremely hard to detect. 


{\em Conclusion.} We have developed a field theory for monopoles in degenerate square ice. In absence of a real monopole interaction the system is a 2D-Coulomb gas where monopoles interact entropically, are always screened, and thus in an unbound, conductive phase. This case has been recently realized  in quantum dots~\cite{king2020quantum}. When the 3D-Coulomb interaction among monopoles is considered, reduced dimensionality prevents full screening, and a dimensional mismatch between Green functions of the Laplace operator drive different effective screening regimes.
 
Our results, obtained via approximations on a simplified mode, invite experimental tests which are however non-trivial: the algebraic behavior can be camouflaged by pseudo-screening if the monopole-monopole interaction is small, and detection might require large real-space characterization. Algebraic correlations might be experimentally detectable in weak singularities near the pinch points, from which the Bjerrum length can be extracted. 

In the future, more precise expressions for various quantities can be computed by Feynman diagram  expansion of ${\cal F}[\phi]$. The spirit of our approach can be applied also to 3D spin ice and to honeycomb/Kagome spin ice~\cite{qi2008direct,Zhang2013}, and can be  extended  to include topological currents beside charges, thus underscoring the gauge-free duality of the square geometry.




\bibliography{library2.bib}{}

\end{document}